\documentclass[
  ,final            % use final for the camera ready runs
  ]
  {aipproc}

\layoutstyle{6x9}

\begin{document}

\title{Gamma-ray Bursts, Classified Physically}


\author{Joshua S. Bloom}{
  address={601 Campbell Hall, Berkeley, CA 94720 USA}
}

\author{Nathaniel R. Butler}{
  address={601 Campbell Hall, Berkeley, CA 94720 USA}
}

\author{Daniel A. Perley}{
  address={601 Campbell Hall, Berkeley, CA 94720 USA}
}

\begin{abstract}
From Galactic binary sources, to extragalactic magnetized neutron stars, to long-duration GRBs without associated supernovae, the types of sources we now believe capable of producing bursts of gamma-rays continues to grow apace. With this emergent diversity comes the recognition that the traditional (and newly formulated) high-energy observables used for identifying sub-classes does not provide an adequate one-to-one mapping to progenitors. The popular classification of some $> 100$ sec duration GRBs as ``short bursts'' is not only an unpalatable retronym and syntactically oxymoronic but highlights the difficultly of using what was once a purely phenomenological classification to encode our understanding of the physics that gives rise to the events. Here we propose a physically based classification scheme designed to coexist with the phenomenological system already in place and argue for its utility and necessity.
\end{abstract}

\maketitle

%%%%%%%%%%%%%%%%%%%%%%%%%%%%%%%%%%%%%%%%%%%%
%% MAINMATTER
%%%%%%%%%%%%%%%%%%%%%%%%%%%%%%%%%%%%%%%%%%%%

%\section{Introduction}

For 30  years since discovery, high-energy observations defined not only the phenomenological class of GRBs but comprised most of the constraints on the physical origin of the events. 
The advent of the afterglow era broadened the scope of this understanding, allowing detailed calorimetry of sub-components that make up the totality of the phenomena: the prompt emission, the blastwave, the trans-relativistic flow, and, in some cases, the supernova component. 
Considering that neutrinos and gravitational waves may be  substantial channels for energy release, we now believe that the $\gamma$-rays of GRBs trace only the tip of the iceberg in the energetics budget (e.g., \cite{wb06}). Classifying and following where the energy isn't can only get us so far in the pursuit to understand the events themselves, the progenitors, and the connection of such events to other explosive phenomena in the universe. A purely phenomenological classification scheme holds some advantage in that it allows quick allocation of resources based on past experience. However, the danger is that such classifications based on the set of the most readily identifiable observable features of an event can inadvertently group heterogeneous progenitor sources into what appears as a homogeneous phenomenological class. Differences from event to event that are both subtle and dramatic can belie vastly different origins.  

\subsection{Phenomenological Classifications of the Past}

Building upon earlier work in the time-domain analysis of GRBs \cite{mg81,ncdt84}, Kouveliotou et al.\ \cite{kmf+93} discovered a bimodality in the duration and spectral hardness plane of GRBs. This work, based on BATSE events, gave rise to the canonical separation\footnote{This dividing line is clearly instrument and bandpass dependent \cite{cbbk06}.} of $t_{90} = 2$ sec for short-hard bursts (SHBs) and long-soft bursts (LSBs). This also gave rise to the long-standing supposition that these two phenomenological classes represented emission from two distinct physical sources. Indeed, in the early days of short-burst discoveries, we advanced that the analogy that ``type Ia supernovae are to core-collapsed supernovae as short-hard bursts are to long-soft bursts'' (``Ia:CC::SHB:LSB''; \cite{bp06}) would be useful in highlighting not only similar environmental observables between the two phenomena (e.g., host galaxy types) but in the drawing out of the physical analogs of the progenitors, particularly degenerate vs.\ non-degenerate.

This otherwise tidy classification scheme --- mapping just two observables to two progenitor classes --- was already challenged on a number of fronts and would be soon challenged with more counter-examples discovered by Swift and the IPN:

\begin{figure}
		\includegraphics[width=.80\textwidth]{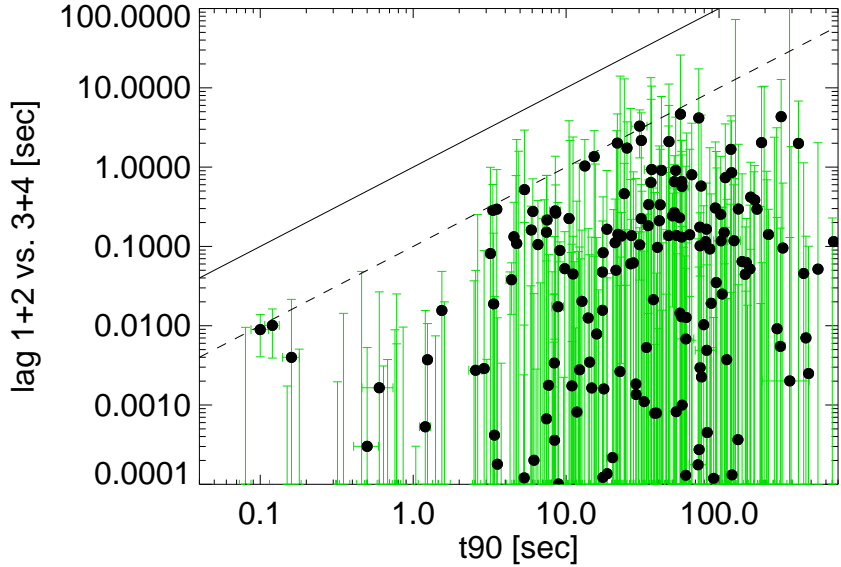}

 	\caption{Demonstration of the Covariance of Lag with Burst Duration ($t_{90}$) for 265 Swift bursts up to and including GRB\,071031. Shown are the inferred lags and associated 2-$\sigma$ uncertainties between BAT channels 1+2 and 3+4, derived from cross-correlation and bootstrap replacement for error analysis.  The data appear constrained by the lag = $0.1 t_{90}$ line (dashed; the solid line is lag = $t_{90}$. Dividing $t_{90}$ by lag, we find that the distribution between the classical ``short bursts'' ($t_{90}$ = 2.0 sec) and ``long bursts'' are indistinguishable: in only 7.2\% of bootstrapped KS trials between the two population would we have noted a $P_{\rm KS} < 0.05\%$. From \cite{cbbk06}, see also \cite{hbn+08}. \label{fig:ks}}
\end{figure}

\begin{itemize} 
	\item {\bf X-ray Flashes (XRFs)}. Technically a class of LSBs, there was never a strong argument made for XRFs simply populating the soft-end continuum in spectral hardness. They might still be shown to arise from a fundamentally different sort of progenitor than the tradition class of LSBs.
	\item {\bf Megaflares from Soft Gamma-Ray Repeaters (SGRs)}. 
	%These most energetic of the Galactic highly-magnetized neutron star events exhibited hard initial spikes, like in SHBs. 
	Tanvir et al.\ \cite{tcl+05} argued that very bright flares could be seen from other galaxies to the point of indistinguishably co-mingling with the ``cosmological'' SHBs in BATSE. Swift SHBs 051103 and 070201 are now identified, based on spatial coincidences, as probable extragalactic magnetars events (see, e.g., \cite{fpa+07,perl08}); without good localizations they would likely have been classified as SHBs.
	\item {\bf Long-Duration Short Bursts}. Events exhibiting short timescale hard-spectrum emission followed by softer and longer emission, sometimes with as much energy as the prompt spike. Here, the $t_{90}$ duration of the event is highly dependent upon the sensitivity of the instrument. Traditional duration analysis at Swift sensitivities placed such events in the LSB category. There were already hints of such events from BATSE \cite{lrg01}.
	\item {\bf Supernova-less Long-Soft Bursts}. Prototypical examples of nearby LSBs without supernovae to deep limits are 060505 and 060614 \cite{fwt+06,gnb+06} but others may also have been seen (e.g., 051109b; \cite{perl08}).
	\item {\bf Long-Soft GRBs from Galactic Binaries}. See Kasliwal et al. \cite{mck+07}.

\end{itemize}	
The addition of light curve {\it lag} between 2 energy ranges was seen as a promising tool to resurrect the observable mappings to a two progenitor class system (see however Figure \ref{fig:ks}). The addition of several more observables, many involving observations at other wavelengths, was introduced \cite{dls+06} to the map a burst probabilistically as belonging to one of two classes (``long population'' or ``short population''). Zhang et al.\ \cite{zzl+07} citing the analogy with supernovae, proposed two phenomenological classes, related to SHBs and LSBs.

Our principal concern is that the LSBs and SHBs (or classes II and I in the Zhang et al.\ prescription), are becoming semantic code within the community for specific physical progenitor models, namely collapsars and binary degenerate mergers. Indeed the careful gerrymandering of event observables into two classes necessarily excludes the diversity of the physical phenomena that give rise to the zoo of high-energy transients. Just as the progenitors that give rise to Type I supernova are a very heterogeneous lot (core-collapsed and WD events), so too are Zhang Type I GRBs.
\begin{figure}[tb]
  \includegraphics[height=0.70\textheight,angle=90]{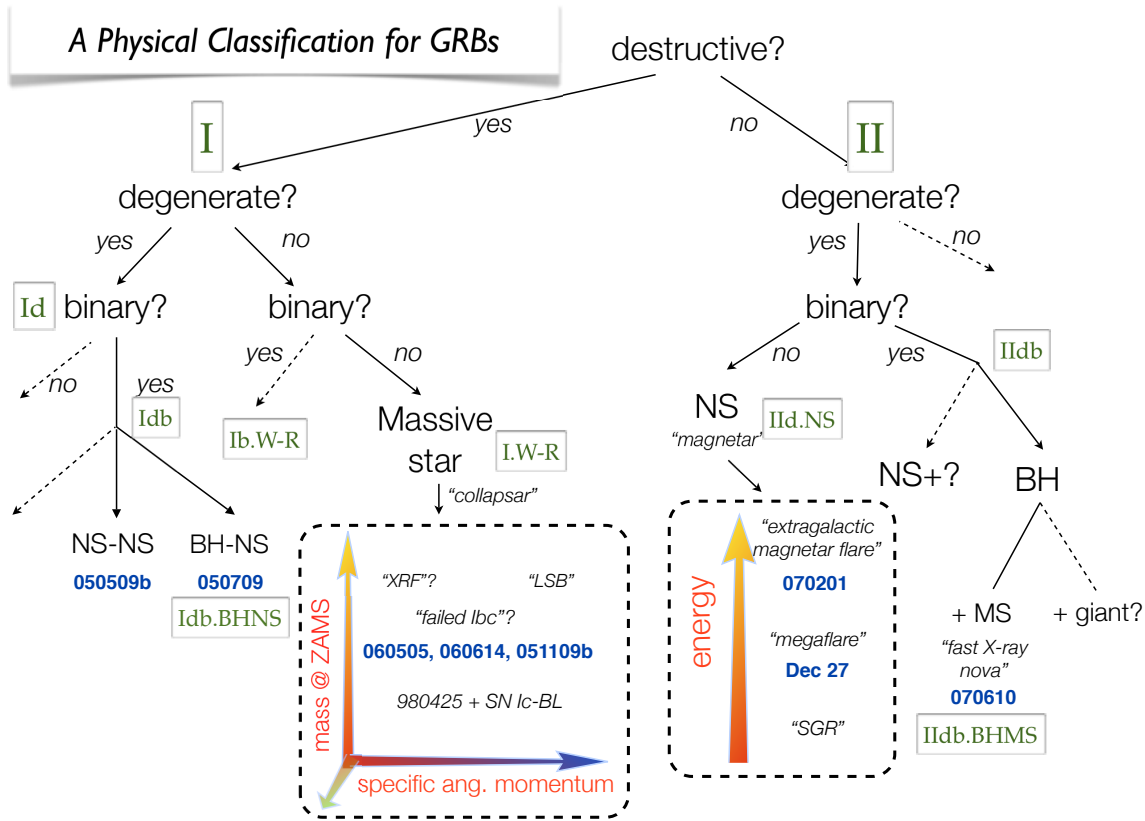}
  \caption{Physical classification scheme based on some current popular progenitor models. Event names in bold-blue are taken to be representative prototypes of the class. Colloquial nomenclature given in quotes while dashed arrow lines indicate tentative or unknown branchings in the decision tree. Axes within the dashed-line boxes are meant to illustrate how a range in a few physical parameters (e.g., energy release, specific angular momentum, ZAMS mass) could give rise to the diversity of the observables within each physical class. The suggested physical classification nomenclature is shown with boxed green lettering.}
\label{fig:classy}
\end{figure}

\section{A Physical Basis for Classifying}

We propose a classification scheme based on the nature of the progenitors and a physical description of the origin of the emission. Figure \ref{fig:classy} illustrates the breakdown of the classification. Progenitor scenarios than cannot repeat either because they are destroyed or fundamentally altered during the event shall be called {\it Type I} sources. {\it Type II} (``non-destructive'') sources are those where the progenitor remains after the event. Systems involving at least one object supported by degeneracy pressure shall be denoted by ``d'', and binary systems where two objects participate substantially in causing the event shall be denoted with a ``b''. For example, a GRB from a degenerate binary merger event comes from a {\it Type Idb} source while an event from an isolated degenerate source that could repeat is said to come from a {\it Type IId} source. Demarcations of the specific progenitors can be added with a period and then in the descending order of mass of each component. A merging black hole--white dwarf system is a {\it Type Idb.BHWD}. Further modification, related to the physical nature of the progenitors pre-explosion (e.g. specific angular momentum), may be captured with another period plus some encoding for the different physical state.
\begin{figure}[t]
  \includegraphics[angle=90,height=0.5\textheight]{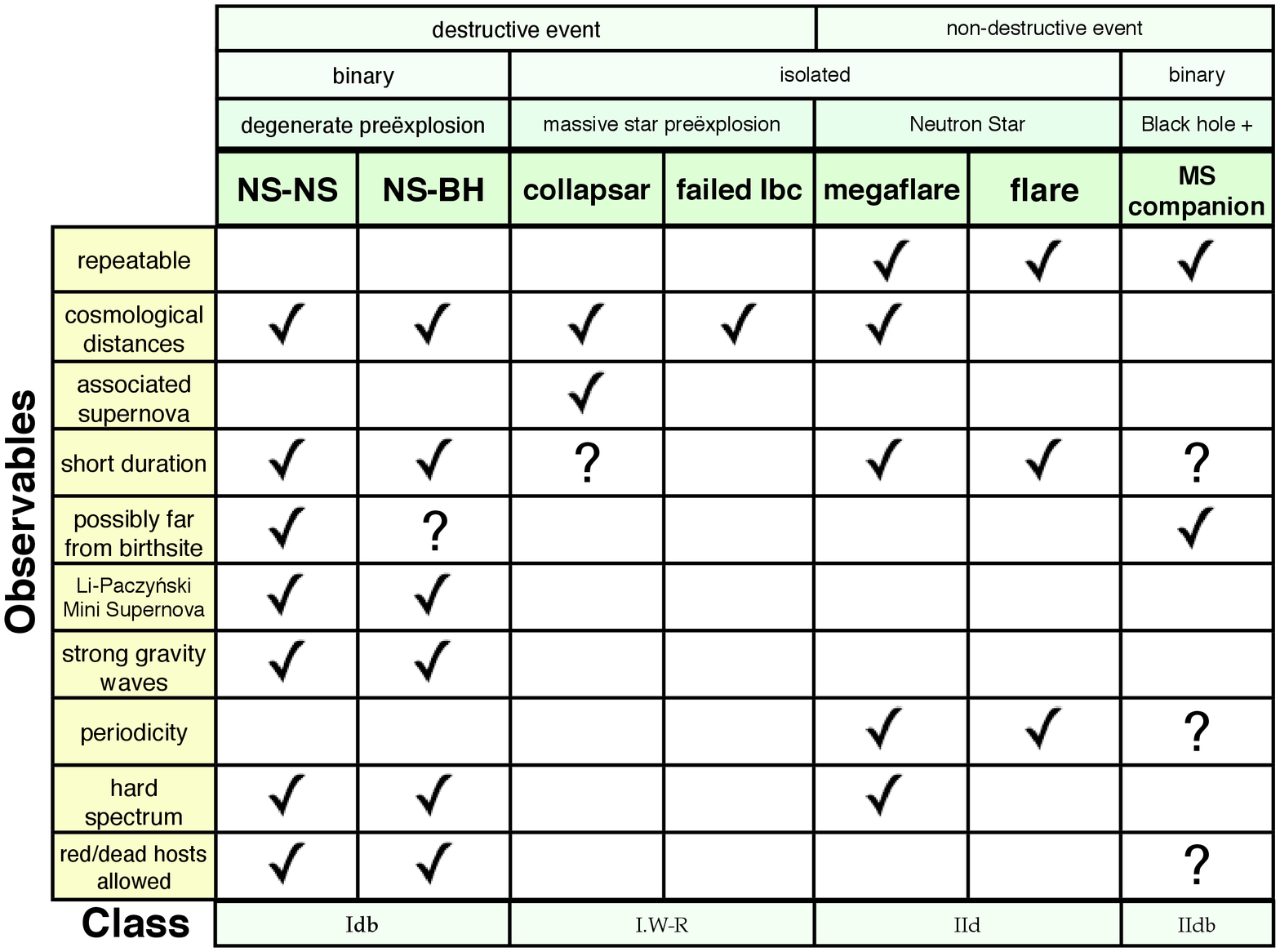}
  \caption{Example mappings of observables to physical classes.}
\label{fig:matrix}
\end{figure}

This nomenclature is attractive because it is a) extensible in obvious ways as new progenitors are proposed and b) simply cannot be ``wrong'' --- only the mapping between the physical sub-class and the range of observables can require modification as the theory evolves. There may never be a GRB from a {\it Type Idb} source, but we know such sources exist in nature. Figure \ref{fig:matrix} highlights the connection of the physical classification to some reasonable statements about observables. It is important to recognize that a ``Short Hard Burst'' may arise from one of many sub-classes of {\it Type Idb} sources, {\it Type I.W-R} (``collapsars''), or a {\it Type IId.NS}. Indeed the phenomenological class of ``SHBs'' could actually be a {\it bona fide} admixture of all three physical classes. Likewise, long-soft bursts (``LSBs'') are likely due to {\it Type I.W-R} and {\it IIdb.BHMS} sources (Kasliwal, this meeting).

We see this physically based classification scheme not as a stark departure from where the field is already heading but as a logical expansion of the descriptive tools we use for further inquiry. We are not advocating for the overthrow of the phenomenological classification of GRBs --- it is clear that rapid identification of observable features has utility --- but with the co-existence of both forms of classification. The advantage here is that just as the physically meaningful set of classifiers does not pre-suppose observbles so too should the phenomenological classification eschew physical preconceptions of the progenitors.

Of course, we are aware that despite the attractiveness of the Shklovskii--da Silva physical classification for supernovae \cite{shk82,das93}, it is the Minkowski-Zwicky phenomenological nomenclature (along with modifications) that has endured. While M-Z may be historically useful, otherwise strange supernovae in the M-Z classification system (e.g., chameleon supernovae, like 2005aj, morphing from IIn$\rightarrow$Ia; Ia supernovae with hydrogen in the spectrum), are trivially explained when viewed from the progenitor formation scenarios and progenitor environments. To be sure, all classification schema that account at a proper depth for both the rich diversity of observables and progenitors will be semantically identical even if syntactically distinct. Ultimately, however, the most useful classification scheme will be one that aids in the most efficient use of scare resources for follow-up observations, to provide the most diverse input to theoretical models.

\begin{theacknowledgments}
  We thank D.\ Kocevski, E. Ramirez-Ruiz, E. Troja, D.\ Poznanski, P.\ Nugent, and E.\ Nakar for lively conversations. We also thank M.\ Galassi, E.\ E.\ Fenimore, and the local organizing committee for a most enjoyable and productive conference. N.R.B.\ is partially supported by the DOE SciDAC grant DE-FC02-06ER41453.
\end{theacknowledgments}

\end{document}